\documentclass{epl}

\newcommand{\Ham}{\mathcal{H}}

\newcommand{\Vc}{V_{\ab{c}}}
\newcommand{\dk}{\upd^3\vect{k}}
\newcommand{\muB}{\mu_{\ab{B}}}

\newcommand{\erfc}{\mathrm{erfc}}
\newcommand{\ED}{E_{\ab{D}}}
\newcommand{\Fe}{\chem{Fe_8}}
\newcommand{\Mn}{\chem{Mn_{12}}-\chem{Ac}}

\newcommand{\ket}[1]{|#1\rangle}

\begin{document}


\title{Dipolar ordering in \Fe{}?}

\author{
  X. Mart\'{\i}nez-Hidalgo\inst{1} \thanks{E-mail:
    \email{xavim@ubxlab.com}}
  \and E. M. Chudnovsky\inst{2} \thanks{E-mail:
    \email{chudnov@lehman.cuny.edu}}
  \and A. Aharony\inst{3} \thanks{E-mail:
    \email{aharony@post.tau.ac.il}}
}
\shortauthor{X. Mart\'{\i}nez-Hidalgo \etal}

\institute{
  \inst{1} Departament de F\'{\i}sica Fonamental, Universitat de
  Barcelona - Diagonal 647, 08028 Barcelona, Spain \\
  \inst{2} Department of Physics and Astronomy, Lehman College, City
  University of New York - Bedford Park Boulevard West, Bronx, New
  York 10468-1589, USA \\
  \inst{3} School of Physics and Astronomy, Raymond and Beverly Sackler 
  Faculty of Exact Sciences, Tel Aviv University - Tel Aviv 69978, Israel
}


\maketitle
\begin{abstract}
  We show that the low-temperature physics of molecular nanomagnets,
  contrary to the prevailing one-molecule picture, must be determined
  by the long-range magnetic ordering due to many-body dipolar
  interactions.  The calculations here performed, using Ewald's
  summation, suggest a ferromagnetic ground state with a Curie
  temperature of about 130~mK.  The energy of this state is quite
  close to those of an antiferromagnetic state and to a glass of
  frozen spin chains.  The latter may be realized at finite
  temperature due to its high entropy.
\end{abstract}
%



In recent years, molecular nanomagnets \Mn{} and \Fe{} of spin 10
\cite{Sessoli,Barra} have been the object of intense theoretical and
experimental studies due to their interesting magnetic properties.
They exhibit spectacular quantum effects, such as regular steps in the
magnetic hysteresis \cite{Friedman,Sangregorio}, quantum interference
effects in the magnetic relaxation \cite{Wernsdorfer}, quantum
avalanches \cite{Paulsen,avalanche}, coherent quantum oscillations of
spin between two classical energy minima \cite{delBarco}, first- and
second-order crossover between thermal and quantum regimes
\cite{Kent}, etc. Most of these effects can be observed in macroscopic
measurements and can be studied by quasi-classical methods, which
places the physics of high-spin molecular clusters in the domain of
macroscopic quantum tunneling \cite{Leggett,book}. A significant
theoretical effort has been mounted to explain the observed phenomena
\cite{theory}. Most of the theoretical works to date treated \Mn{} and
\Fe{} crystals as consisting of non-interacting spin-10 clusters,
addressing basically the question of transitions between spin levels
of individual clusters due to internal and external fields, phonons,
and nuclear spins. More recently Prokof'ev and Stamp \cite{PS}
suggested that in certain experiments the dipolar interactions between
clusters can be responsible for the relaxation law (see also
\cite{comment,miya00}). This suggestion has been confirmed by computer
simulations \cite{PS,Villain} and by the experimental evidence of
memory effects \cite{wernHD,memory} recently observed in both \Mn{}
and \Fe{}.

The question addressed by us is this: can dipolar interactions induce
magnetic ordering in \Mn{} and \Fe{}? If this was true, the
low temperature physics of molecular nanomagnets would be the physics
of magnetic order in a many-body system, which would be more
interesting than the current spin-10 physics. 
It is important to emphasize that the possibility of ordering has been
overlooked in previous works studying, e.g., the relaxation of these
systems.  For instance, the hole-digging method~\cite{wernHD} can be
regarded, in the light of this paper, as due to irreversibility and
memory effects in a spin-glass phase (a well-known
effect~\cite{erikvincent}).
The subject of magnetic ordering induced solely by dipolar
interactions is not new
\cite{lutttisz46,cohekeff57,niem72,aharfish73}. Experiments
established magnetic order in dilute rare earth insulating
compounds\cite{whiroscor}. Theoretical studies performed to date were
restricted mainly to cubic lattices.  The main theoretical
result\cite{lutttisz46} is that the lowest energy configuration for a
simple cubic (sc) lattice corresponds to an antiferromagnetic
configuration (with ordering vector $\vect{k}=(0.5, 0.5, 0)$), while
in the body-centered (bcc) and the face-centered (fcc) cubic case, the
ferromagnetic configuration is energetically favorable.

Both \Mn{} and \Fe{} have strong easy-axis anisotropy. The
$H=0$ problem is then the one of an Ising system interacting via
magnetic dipole forces. There is a significant difference between the
two systems, however. Firstly, in \Mn{} the terms breaking the
uniaxial magnetic symmetry at $H=0$ are weak. Consequently, at low
temperature ($T<1$~K), the spins of \Mn{} molecules are blocked
along the anisotropy directions chosen by the history of the cooling
process. In addition, the hyperfine fields in \Mn{} are as strong
as the dipole fields but random from one \Mn{} cluster to another.
Consequently, a highly metastable spin-glass order, if any, should be
expected in \Mn{} as temperature is lowered.

In \Fe{} the terms violating the uniaxial symmetry are large, leading
to thousands of quantum transitions per second between the equivalent
easy-axis directions at $T=0$ \cite{Wernsdorfer}. Also, \Fe{} has
smaller lattice parameters
($a=10.522(7)$~\AA, $b=14.05(1)$~\AA, $c=15.00(1)$~\AA,
$\alpha=89.90(6)^\circ$, $\beta=109.65(5)^\circ$,
$\gamma=109.27(6)^\circ$~\cite{wiegetal84})%
, making dipole interactions more important,
while hyperfine interactions in \Fe{} are very weak. 
The \Fe{} cluster is chemically constituted by an octameric
Fe$^{\mathrm{III}}$ cation, in which six of the Fe$^{\mathrm{III}}$ atoms
coordinate with a cyclic amina.  The cyclic amina ligands form the
``surface'' of the complex cation, effectively isolating the
Fe$^{\mathrm{III}}$ atoms from the surrounding (more details on the
chemical structure of \Fe{} can be found in Ref.~\cite{wiegetal84}) .
Therefore, other interactions between \Fe{} clusters, such as exchange
or super exchange, are also negligible.  Thus, in \Fe{} the low
temperature magnetically ordered ground state due to dipole
interactions must be simpler and much easier to achieve than in \Mn{}.
The estimate of the temperature at which dipolar interactions might
begin to be of importance can be obtained from the order of magnitude
of the dipolar energy.  One gets $T_{\ab{dip}}\approx \ED =
(g\muB{}S)^2/\Vc \approx 130$~mK, where $\muB$ is the Bohr magneton,
$g\approx 2$ is the gyromagnetic ratio\cite{Barra,barr00,caci98} and
$\Vc\approx 1956$~\AA$^3$ is the volume of the unit
cell\cite{wiegetal84} of \Fe{}.
Note that if the ordering is ferromagnetic (as will be shown below),
deviations of the actual critical temperature from this mean-field
estimate should not be too large, as the upper critical dimension for
an Ising dipolar ferromagnet is $d=3$~\cite{aharfish73}.
In this work we shall find the
classical ground state of \Fe{} due to dipole interactions,
assuming that the only role of the terms in the Hamiltonian which are
responsible for tunneling between individual easy directions, is to
help to achieve that ground state.

To find, through direct minimization, the lowest energy configuration,
one can study the Fourier transform of the dipolar interaction.  Here
we are assuming for simplicity that each \Fe{} cluster is
characterized by a single dipole moment located at the center of the
cluster.  Given that magnetic moments of eight iron ions belonging to
an \Fe{} cluster are separated by an appreciable distance, this, of
course, is a crude approximation. We believe, however, that it catches
the physics of the problem.

The Hamiltonian we consider for the crystal, assuming that the
external field is zero, is
\begin{equation}
  \label{eq:ham0}
  \Ham = \Ham_{\ab{dipole}} + \Ham_{\ab{single-ion}} \; ,
\end{equation}
where
\begin{equation}
  \label{eq:hamdip}
  \Ham_{\ab{dipole}} = -\frac{1}{2}\sum_{i\neq j} D_{\alpha\beta,ij} 
    S^{\alpha}_{i} S^{\beta}_{j}
\end{equation}
and 
\begin{equation}
  \label{eq:hamsingle}
  \Ham_{\ab{single-ion}} = \sum_{i} -K( {S^{z}_{i}}^2
  + b {S^{x}_{i}}^2 ) \;\;.
\end{equation}
For \Fe{}, one has $K\approx 230$~mK, $b\approx 0.4$ and $\ED \approx
130$~mK~\cite{Barra,barr00,caci98}.  We will consider the limit in
which the spins align themselves along the easy axis (the $z$ axis in
our notation), which we take to be approximately parallel to the
crystallographic axis $a$\cite{barr00}.  Thus we can project each spin
onto the basis $\ket{S_z=\pm{}S}$.  Note that the energy difference
for a small angle rotation from $\ket{S_z=S}$ to $\ket{S_z=S-1}$ is
$\Delta E \approx K(2S-1) \approx 5$~K~$\gg \ED$ in \Fe{}, so this
approximation is quite accurate in the relevant range of temperature.
Taking into account these considerations, we obtain an Ising-like
array of spins with the effective Hamiltonian
\begin{equation}
  \label{eq:hamising}
  \Ham = -\frac{\ED}{2}\sum_{i\neq j} \bar{D}_{zz,ij} \tau^z_i \tau^z_j
    + \sum_{i} \Delta \tau^x_i \;,
\end{equation}
where $\bar{D}_{zz,ij}=[3\cos^2\theta_{ij}-1]\Vc/r_{ij}^3$,
$\theta_{ij}$ is the angle between $\vect{r}_{ij}$ and the easy
$z$-axis and $\Delta$ is the tunnel splitting responsible of tunneling
between the $\ket{S_z=\pm{}S}$ states.  Here $\tau^z_i$ is a Pauli
matrix for the site $i$.  If the main source of tunneling is the
${S^x_i}^2$ term, $\Delta$ is of order $K
b^{S}$\cite{gara00}%
(for \Fe{} a typical value is $\Delta \approx 10^{-7}$~K $\ll E_D$)%
.  This Hamiltonian describes the "quantum spin glass"
problem\cite{thilhuse95,cuglloza98}.  From now on, we will restrict
our considerations to the {\it classical\/} approximation, i.e., 
$S\rightarrow\infty$.  We will accordingly neglect the tunneling
term%
\footnote{ The tunneling term is necessary to obtain the true
  quantum ground state.  Even at finite temperatures, it induces
  fluctuations that might destroy the order found here.  
}%
and regard ${\tau^z_i}$ as a number.  Thus, the problem reduces to
finding the configuration $\{\tau^z_i\}$ which minimizes
\begin{equation}
  \label{eq:hamclass}
  \Ham = -\frac{\ED}{2}\sum_{i\neq j} \bar{D}_{zz,ij} \tau^z_i \tau^z_j \;,
\end{equation}
with the restriction
\begin{equation}
  \label{eq:ising}
  \tau^z_i = \pm 1, \qquad \forall i \;.
\end{equation}
We consider the Hamiltonian in the momentum space
\begin{equation}
  \label{eq:hamk}
  \Ham/\ED = -\frac{\Vc}{(2\pi)^3} \int\dk{}\;\tau^z(\vect{k}) 
             \bar{D}_{zz}(\vect{k}) \tau^z(-\vect{k})  \;,
\end{equation}
where the integration is performed over the first Brillouin zone and
$\tau^z(\vect{k})$ and $\bar{D}_{zz}(\vect{k})$ are the Fourier
transforms of $\tau^z_i$ and $\bar{D}_{zz,ij}$, respectively, given by
\begin{eqnarray}
  \label{eq:tauk}
  \tau^z(\vect{k}) = \sum_i \tau^z_i e^{-i \vect{k}\cdot\vect{r}_i}
  \;\;, \\
  \label{eq:dk}
  \bar{D}_{zz}(\vect{k}) = \sum_{i\neq 0} \bar{D}_{zz,i0} \,
       e^{i \vect{k}\cdot\vect{r}_i}  \;\; .
\end{eqnarray}
Here we have made use of the translational invariance of
$\bar{D}_{zz,ij}$.  The sum in Eq.~(\ref{eq:dk}) converges very
slowly.  In fact, it is non-analytical as $\vect{k} \rightarrow 0$,
this being a manifestation of the shape dependence of the energy in
the presence of a net magnetization.  This sum has been computed using
Ewald's method\cite{aharfish73,ewal21,bowdclar81}.  In this method,
the Fourier transform of the dipolar tensor is given by
\begin{eqnarray}
  \label{eq:ewald}
  \bar{D}_{\alpha \beta}(\vect{k}) 
     &=& -4\pi \frac{k_\alpha k_\beta}{k^2} 
          \exp({-\pi^2 k^2 / R^2 }) \nonumber\\
       &&- 4\pi {\sum_{\vect{G}_i}}' \frac{b_{i,\alpha} 
           b_{i,\beta}}{b_i^2} \exp({-\pi^2 b_i^2 / R^2}) \nonumber\\
       &&+ \sum_{\vect{r}_i} \left\{ \frac{2}{\sqrt{\pi}} x_i e^{-x_i^2} 
           \left[ (3+2x_i^2) \frac{r_{i,\alpha} r_{i,\beta}}{r_i^2} 
             - g_{\alpha\beta} \right] \right. \nonumber\\
       &&\;\; \left. + \frac{\Vc}{r_i^3} \left[ 3 \frac{r_{i,\alpha} 
             r_{i,\beta}}{r_i^2} - g_{\alpha\beta} \right] \erfc(x_i) 
             \right\} \exp({i 2\pi k_\mu r_i^\mu}) \nonumber\\
       &&+ \frac{4 R^3 \Vc}{3\sqrt{\pi}} g_{\alpha\beta}  \;\;,
\end{eqnarray}
where $\vect{G}_i$ runs over the reciprocal lattice (the prime
indicates that we exclude the term for $\vect{G}_i=(0,0,0)$),
$\vect{r}_i$ runs over the Bravais lattice, $x_i= Rr_i$, $\vect{b}_i =
\vect{G}_i + \vect{k}$ and $g_{\alpha\beta}$ is the metric associated
with the unit cell basis.  Note that the basis of the reciprocal
lattice used to expand the reciprocal space vectors in components is
the dual basis ${\vect{e}^\alpha}$ of the real space unit basis
$\vect{e}_\beta$, defined by $\vect{e}^\alpha(\vect{e}_\beta) =
\delta^\alpha_\beta$ (and not by the more conventional relation
$\vect{e}^\alpha(\vect{e}_\beta) = 2\pi\delta^\alpha_\beta$).  $R$ is a real
parameter which controls the convergence of the series (the value of
$\bar{D}$ is independent of it).  For a cubic lattice, the optimal value is
$R=2.0/a$, whereas for \Fe{} it has been found that convergence is
fastest for both the reciprocal and the real space sum using
$R=1.8/\Vc^{1/3}$.

Now it must be noted that one cannot realize spin structures for
arbitrary $\vect{k}$ with Ising spins.  Indeed, in order to satisfy
restriction~(\ref{eq:ising}), $\vect{k}$ must be either in the center
of the Brillouin zone (ferromagnetic arrangement) or at its boundary,
i.e., we must have $k_\mu=0,1/2$, with $\mu=1,2,3$.  Moreover, the
dipolar interaction energetically favors the ferromagnetic alignment
of spins in chains along the $a$ axis.  Thus, we will restrict our
consideration to the case $k_1=0$ (configurations with $k_1=1/2$ have
been found to have much higher energy, as expected).  The distinct
configurations we need to consider are
\begin{eqnarray}
  \label{eq:ks}
  &\vect{k}_0 = (0,0,0)\;, \qquad \vect{k}_1=(0,0,1/2) \;, \nonumber\\
  &\vect{k}_2 = (0,1/2,0) \;, \qquad \vect{k}_3=(0,1/2,1/2) \;.
\end{eqnarray}
Here $\vect{k}_0$ corresponds to a ferromagnetic arrangement, and
$\vect{k}_1$, $\vect{k}_2$ and $\vect{k}_3$ correspond to
antiferromagnetic arrangements with planes of alternating
magnetization ($[001]$, $[010]$ and $[011]$-planes, respectively).
The energy per spin of these configurations can be computed using
Eq.~(\ref{eq:ewald}) and the relation $E = -(1/2)
\bar{D}_{zz}(\vect{k}) \ED$.  One obtains
\begin{eqnarray}
  \label{eq:energies}
  &E_0 = -4.10 \,\ED \;, \qquad E_1 = -4.05 \,\ED \;, \nonumber\\
  &E_2 = -4.02 \,\ED \;, \qquad E_3 = -3.97 \,\ED \;.
\end{eqnarray}
Therefore, we find that the lowest energy configuration corresponds to
ferromagnetic alignment of the spins, with an energy per spin
$E_0\approx-530$~mK (note that due to the smallness of the energy
differences, unaccounted effects may, in principle, shift the ground
state; this will be discussed elsewhere).  In evaluating $E_0$, one
must note that the non-analyticity of the first term in
Eq.~(\ref{eq:ewald}) can be attributed to shape-dependent
demagnetizing effects, as discussed above.  We can define the
demagnetizing factor as
\begin{equation}
  \label{eq:demagnet}
  L = 4\pi \lim_{\vect{k}\rightarrow (0,0,0)} \left(\frac{k_z}{k}\right)^2 \;.
\end{equation}
Then, one must take $L=0$ for needle-shaped samples (or for systems
with periodic boundary conditions without any free surfaces), which is
the situation we have assumed in computing $E_0$.  Nevertheless, the
bulk free energy of the system should be shape independent in the
absence of an external field (see, e.g., Refs.~\cite{kitt51,grif68}).
Physically, for other sample shapes the demagnetizing energy is
reduced due to the formation of domains in the ferromagnetic state.
In this case, though, we should include a domain wall energy
contribution.  However, this contribution can be shown to be
negligible in the thermodynamic limit due to the macroscopic size of
the domains\cite{mitsfuru53}, and hence $E_0$ should be shape
independent.

Besides these configurations, one must note that no general arguments
exist which would rule out the existence of lower lying ground-states
with structures having larger unit cells (such as arrangements where
two planes of ferromagnetically aligned up-spins alternate with two
planes of down spins).  However, numerical investigation of
$\bar{D}_{zz}(\vect{k})$ by standard optimization methods (e.g., amoeba
method), seems to indicate that $\vect{k}=\vect{k}_0$ is the global
maximum of $\bar{D}_{zz}$.  As any configuration can be expanded in
the ``plane wave'' $k$-basis, we conclude that the ferromagnetic
arrangement is, effectively, the lowest energy configuration.

Particularly interesting and physically relevant (because of its large
entropy) is also a ``spin glass''-like phase, where chains of
ferromagnetically aligned up-spins in the $a$-direction alternate with
chains of down-spins, such that the spin arrangement in the $bc$-plane
is completely random.  The energy of this phase is given by
\begin{equation}
  \label{eq:sg}
  E_{\ab{SG}}/\ED = -\frac{1}{2}\sum_{i(\neq j) , \;
    \vect{r}_{ij}\parallel a} \frac{2 \Vc}{r_{ij}^3} 
     = -2\zeta(3) \frac{\Vc}{a^3} \approx -4.04 \;,
\end{equation}
where $\zeta(p)=\sum_{n=1}^{\infty}n^{-p}$ is the Riemann
zeta-function.

In recent work\cite{fern00}, it has been claimed that the easy axis is
parallel to the crystallographic $b$ axis.  For the sake of
completeness and in order to compare the results here obtained with
those presented in Ref.~\cite{fern00}, we have considered alternative
orientations of the easy axis.  Taking the easy axis parallel to the
crystallographic $b$ axis, one obtains that the lowest energy
configuration corresponds to $\vect{k}_1=(0,0,1/2)$, in agreement with
MC simulation\cite{fern00}.  The energy here computed is
$E_1/\ED=-1.78$, which differs by less than 10\% from the value quoted
in Ref.~\cite{fern00}.  This small difference could be attributed to
finite-size effects and the truncation of the interaction used in MC
simulations.  In the case that the easy axis is taken parallel to the
$c$ axis, the ground state configuration is $\vect{k}_3=(1/2,1/2,0)$,
with energy $E_3/\ED=-1.55$.

Summarizing, the present work suggests the onset of the ferromagnetic
ordering in \Fe{} due to dipolar interactions below $T\approx 130$~mK.
The effect should exist in magnetic fields below the dipole field for
which we obtain $H = \bar{D}_{zz}(\vect{k}_0) g\muB S/\Vc \approx
77$~mT.  It can be observed by conventional methods.
The most direct method would be the observation of a different
relaxation dynamics above and below $T_{\ab{c}}$ for the sample
initially magnetized along the magnetic anisotropy axis. After the
field is turned off, such a sample should relax towards lower
magnetization above $T_{\ab{c}}$ and towards higher magnetization
below $T_{\ab{c}}$.  Measurements of the specific heat of \Fe{} down
to very low temperatures would be helpful as well.  However, achieving
the ferromagnetic phase might be difficult, due to its low entropy and
the presence of high anisotropy barriers, and experiments might probe
a spin-glass ordering, instead.  In that case one should look for
memory effects. In fact, aging and rejuvenation phenomena observed at
mK temperatures in \Mn{} and \Fe{}\cite{wernHD,memory} are very
similar to those found in spin-glasses and disordered
ferromagnets\cite{erikvincent}.
Another interesting problem, not addressed by this work, is the
spectrum of collective excitations in the magnetically ordered phase
of \Fe{}.  The presence of the tunneling term in the Hamiltonian,
suggests the possibility of a spin-flip excitation (``tunnelon'').
This and other questions will be studied elsewhere.

We are grateful to Wolfgang Wernsdorfer for discussion and for
bringing our attention to Ref.~\cite{barr00}.  XMH acknowledges
support from the Generalitat de Catalunya. EMC acknowledges NSF grant
No. 9978882. This work had begun during the authors' stay at the
Institute of Theoretical Physics in Santa Barbara, within the Program
on Magnetic Phenomena in Novel Materials and Geometries.


\end{document}